# How memories are stored in the brain – the declarative memory model

Jie Zhang

March 2024

## Abstract

The ability to form memories is a basic feature of learning and accumulating knowledge. But where is memory information stored in the brain? Within the scientific research community, it is generally believed that memory information is stored in the synapse. However, this widely accepted dogma has been challenged by more and more evidence in recent years. In this paper, I will present a memory model – the neural circuit construction theory - created from a perspective that may completely subvert people's traditional thinking. Instead of looking for whether memories are stored inside neurons or in the synapses, the theory proposes that the human brain does not have any "container-like" memory storages, rather human memory is "hardwired" through "live wiring". The model consists of two parts: the semantic memory model and the episodic memory model.

Today, brain-like, and brain-inspired artificial intelligence are very popular research topics. To design human-like intelligent machines, we must first understand how the human brain works.

The operation of the human brain differs significantly from that of a computer. In the case of a computer, saving an image requires encoding it into a binary format understandable by the machine. Long-term memory (LTM) in computers is typically stored in container-like memory units. When a memory needs to be retrieved later, it must be fetched from LTM and decoded back into its original form.

In contrast, human information processing begins with input from sensory organs that convert physical stimuli such as touch, heat, sound waves, or light photons into electrochemical signals. These signals (an image for example) are then transmitted to the brain's visual center in the form of neural spikes. At this stage, the image is perceived by the mind's eye (the central executive center of vision) in a high image resolution. By focusing, the brain projects the information that needs to be stored to another brain mechanism. The image resolution under this mechanism is much lower than the former. We will refer to this brain mechanism as the saving/retrieving center or S/R center for brevity. During this process, all the information received by the S/R center consists of neural spikes. If these neural spikes can be stored and regenerated later as needed, they constitute human memory. This means that the brain does not need to retain the original sensory input (in this case, the image), but only the neural spikes that the S/R center received (figure 1).

This distinctive characteristic indicates that human memories are formed fundamentally differently compared to contemporary computers. Therefore, I can boldly assert that 1) human memory does not undergo an encoding/decoding process; 2) the human brain does not need container-like storage units; 3) the human brain does not store knowledge in the form of weights; 4) the human brain does not have software-like memory, such as algorithms.

Then, how does the brain store information?

**Semantic memory model**

Hypothesis: Human semantic memory is created by building neural circuits through livewiring in the cerebral cortex.

To create a semantic memory model of human memory, I will make the following assumptions as the baseline.

1). Unlike the artificial neural network, not all the neurons are connected to each other in a human brain. They make connections only when needed. Neural connections in the human brain also do not have synaptic weight. They are either successfully connected or not successfully connected.

2). Human connectome, a comprehensive map of neural connections in the brain, can be divided into two parts: the pre-connected part, and the to-be-connected part. All long-range structural connections are assumed pre-connected before birth. While the most of neocortex is only partially connected and considered as the to-be-connected part.

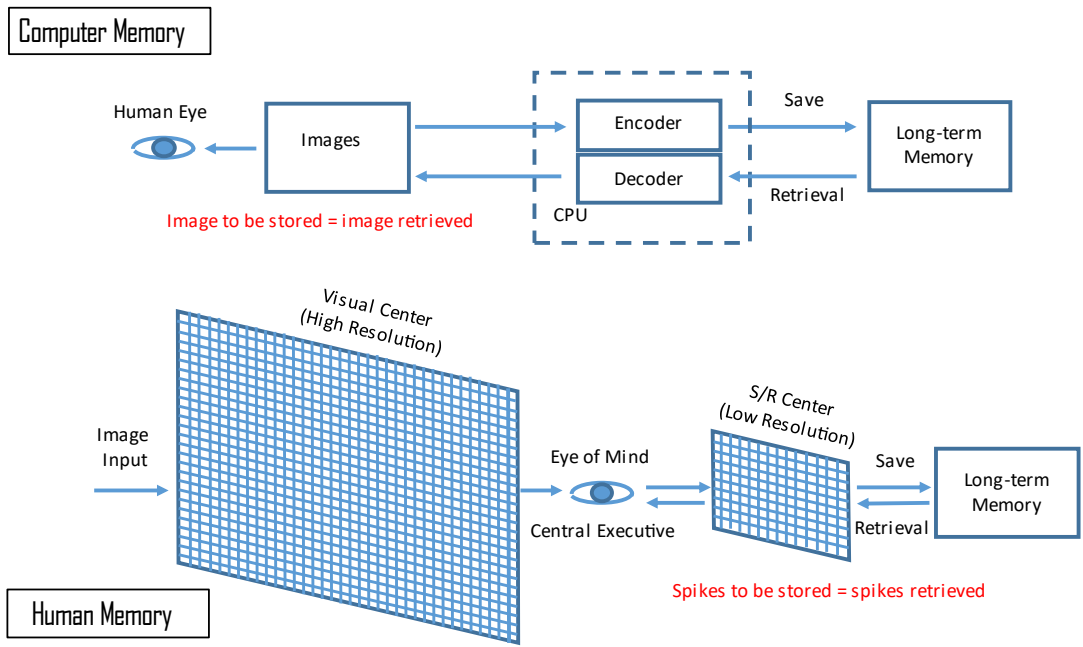

Figure 1. Computer Memory vs Human Memory

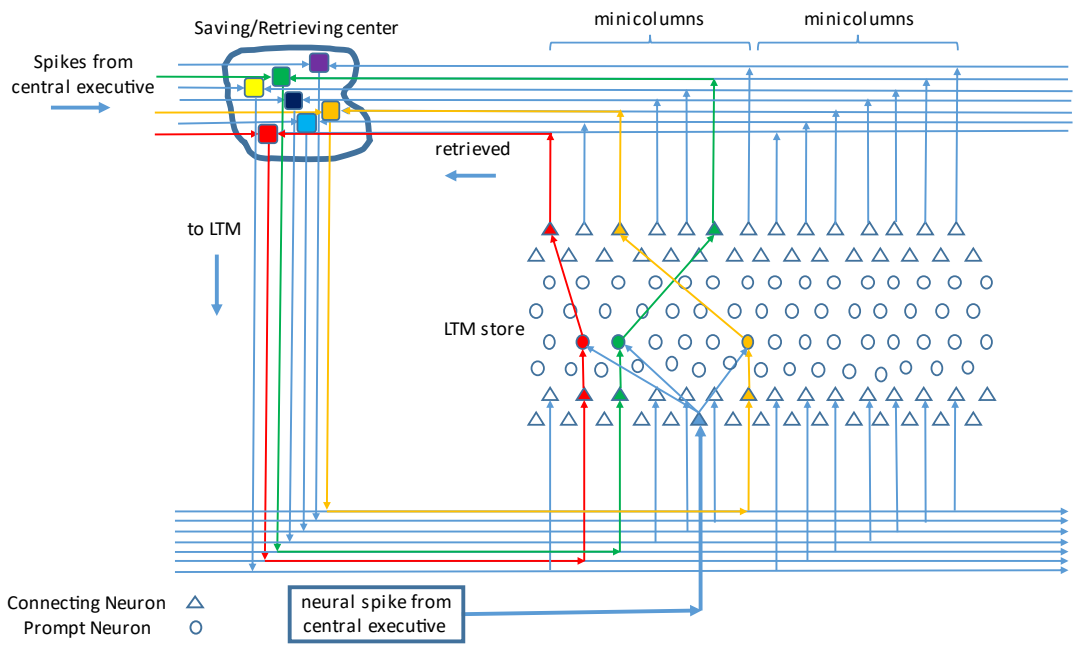

Figure 2. Semantic memory model.

3). As the Nobel laureate Francis Crick put it, we are "nothing but a pack of neurons." From this point of view, I will consider every brain component is formed by "a pack of neurons". Thus, when I say information is sent from component A to component B, I mean that some neurons in pack A send a stimulation signal to activate some neurons in pack B.

4). A human brain has about 86 billion neurons, among them, about 16 billion in the cerebral cortex. These neurons can be divided into two groups: excitatory and inhibitory. The excitatory neurons can be further broken down into two groups: the connecting neurons and the prompt neurons. Therefore, human semantic memory mainly consists of 3 types of neurons: connecting neurons (excitatory), prompt neurons (excitatory), and inhibitory neurons. The connecting neuron's (c-neuron's) function is to build circuits by making connections through axon growth, especially long distance. The prompt neuron's (p-neuron's) function is to lock in on the target neuron and guide the c-neurons to build the correct circuit. The inhibitory neurons' function is to regulate the activity of the circuit. In the cerebral cortex, I assume the stellate (granule) neurons play as the p-neurons, and the rest of the other kinds of excitatory neurons, such as pyramidal neurons, etc. are the c-neurons.

5). The neocortex is made up of six layers, labeled from the outermost inwards, I to VI. The neocortex is also being organized radially into cortical columns. I assume that the outer part of several layers (I to IV) is mainly used as a semantic LTM store, and the inner part of several layers (IV to VI) is mainly used as voluntary motor control. The formation of cortical columns can be explained by the number of neurons in the S/R center. For example, if we assume the visual S/R center contains N neurons, then the visual cortex will be assumed to have N pre-connected neural pathways to each cortical column (See Figure 2). That is to say that the size of the cortical column in the visual cortex is related to the number of neurons in the visual S/R center.

Now, we are ready to model the formation of the semantic memory. I will use a very simple example to illustrate how to build these neural circuits. Let's start with the newborn brain. To make the analysis easier, I assume that all pre-build pathways have been completed, but all to-be-build pathways have not yet been built. In other words, the brain is now in a blank slate (tabula rasa) state. Now, suppose the brain receives its very first visual sensory input: the letter "E" for example. This image of E will be perceived by the eye and converted into neural spikes. Under the stimulation of these neural spikes, some neurons in the visual conscious center will be activated. Although at this stage, the conscious center (our mind's eye) can "see" image E, with a blank LTM, the brain cannot interpret what it perceives. However, with focused attention, the high-resolution E will be projected to the S/R center and stored as a lower-resolution image. Now, the biggest problem here is how to save the image.

Assuming that the S/R center is composed of N neurons, and the image of the letter E activates n of them, then creating the memory of the letter E is to create a neural circuit. Once this circuit is activated, the same neural spike n can be produced. Figure 2 illustrates a possible mechanism for how to create novel semantic memory. In Figure 2, we assume that N=7, and n=3 (represented by red, green, and orange). That is, the S/R center receives a request to create a circuit, and when the circuit is excited, it will generate spikes to stimulate all three neurons (red, green, and orange) simultaneously. We refer to these three neurons in the S/R center as master neurons. Building the circuit consists of 3 steps. 1) follow the pre-established pathway and find 3 free p-neurons in a cortical column. Assign the first of the p-neurons to connect the red neuron in the S/R center, the second p-neuron to connect the green neuron in the S/R center, and the last p-neuron to connect the orange neuron in the S/R center. 2) some neighboring c-neurons will join the circuit-building process to help close the loop. The axons of these c-neurons are directed to grow toward their corresponding target master neurons. The connection rules are a) p-neurons that fire together will be connected together; b) assigned p-neurons will be connected to their corresponding master neurons. 3) since axon growth is time-consuming, it is reasonable to assume that the guidance of these p-neurons toward their targets will demise over time. Therefore, creating a complete circuit requires a lot of repetitive learning trials. Once completed, the semantic memory created can give us i) recognition and familiarity when we see the letter E in the future, and ii) retrieval of the letter E by stimulating this circuit.

Constructed through live wiring, human semantic memory will have the following properties:
- Created from scratch.
- Hard wiring - permanent memory
- Need a lot of repeated learning trials to complete a circuit - slow to build/slow to learn.
- Fast to retrieve, but rigid and inflexible.
- Semantic memory is the building blocks of declarative memory.

**Episodic memory model**

Hypothesis: Human episodic memory is created through the hippocampus to establish temporary connections between existing semantic memories. Traditionally, episodic memory has been defined as a memory involving personal experiences that occurred at specific times and places in the past. This definition is just an extreme case of the definition given in this article.

While the advantages of semantic memory are obvious, its disadvantages can also be devastating. For example, we need the ability to learn quickly, the ability to retrieve memories flexibly, and the ability to forget after remembering for a while. We don't want all memories to be permanent. Some memories are no longer needed after use. For example, we need to remember where we parked our car at the shopping center while we were shopping (fast learning). However, we usually don't need to remember it anymore after we drive away from the shopping center. As for retrieval flexibility, assuming I have only learned semantic memory, and have mastered the following sentence: "Paris is the capital of France". If you asked me to finish the sentence "Paris is the capital of _", I would not hesitate to give you the answer: "France". But if you ask me, "Where is the capital of France?", I will not know how to answer. Therefore, to overcome all the above problems, the brain needs a different structure for episodic memory learning.

Imagine a brain center where all semantic memory storage units (visual, auditory, etc.) are brought together through pre-wired neural networks. They are all pre-connected to each other in the center. The "opening" and "closing" of these connections in the center are controlled by inhibitory neurons. Usually, these connections are in a "closed" state. In this way, the center can immediately establish a temporary circuit between any two semantic memories. Figure 3 illustrates a very simplified brain mechanism with these characteristics. In Fig. 3, the two minicolumns from cortical area A have been preconnected to the two minicolumns from cortical area B. But all these interconnections (the red lines in the figure) are usually blocked by inhibitory neurons (the green perpendicular signs in the figure). The function of these inhibitory neurons is to control the blocking and unblocking of the interconnections. These inhibitory neurons are usually in an "on" state, which causes the interconnections to be blocked. However, when an inhibitory neuron is in an "off" state, the corresponding interconnection becomes unblocked. In this case, the two semantic memories connected by this interconnection become associated. This brain center is what we call the hippocampus.

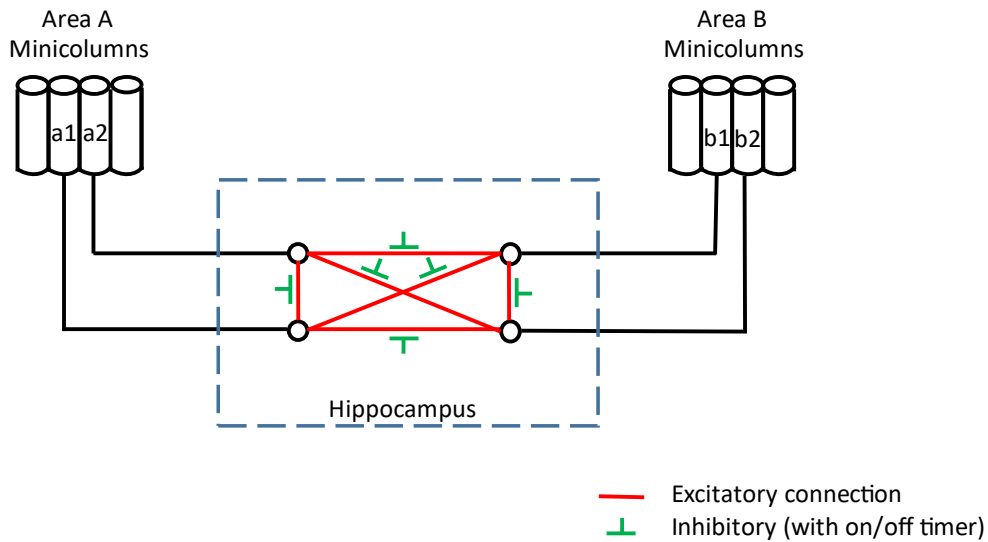

Figure3. Episodic memory model

Since all these interconnections occur between semantic memories, episodic memory is structurally based on semantic memory. The role of inhibitory neurons is to act as switches to control the opening and closing of these links and work in a timer-like manner. Here, the connectivity weakens over time, which can be described by the Ebbinghaus forgetting curve.

Although the hippocampal mechanism does not directly create semantic memories, it will help guide the axons of c-neurons to grow toward their targets when these associations are reactivated especially during sleep. Without the help of the hippocampus, the creation of new semantic memories will be greatly prolonged.

Episodic memory is a kind of long-term declarative memory, which forms a more complex type of memory by associating existing semantic memories. Therefore, semantic memory is the cornerstone of episodic memory. Without semantic memory, there will be no episodic memory. In theory, we must have at least two semantic memories to form an episodic memory. This is why episodic memory does not appear until around 3 to 4 years of age after the baby has created a certain amount of semantic memory. Generally speaking, we cannot judge whether a piece of memory is a semantic memory or an episodic memory, based solely on the content of the memory itself. Just because a memory contains "what", "where", and "when" doesn't necessarily make it an episodic memory, because for any content if we recite it long enough, it becomes a semantic memory. Human declarative memory is not a function of the state but a function of the path through which that memory is retrieved.

The human episodic memory will have the following properties:

- With preconnected pathways, episodic memory can be learned in a single trial. Thus, it belongs to fast learning.
- The association created by the Hippocampus mechanism is temporary. The strength of association decreases with time and can be best described by the Ebbinghaus forgetting curve.
- Associations make human declarative memory much more flexible to retrieve.
- The way episodic memory is formed makes human memory have the ability to forget.

## Summary

The neural circuit construction theory proposes that human semantic memory is created through "livewiring" and is the cornerstone of declarative memory, while human episodic memory is formed by associating semantic memories through the hippocampus. This dual memory structure can lead to a series of dual characteristics of human behavior, as described by Daniel Kahneman in the book "Thinking, Fast and Slow". A memory system created in this way will make any attempt to upload and download human memory impossible.